\documentclass[pra,twocolumn,showpacs,preprintnumbers,amsmath,amssymb]{revtex4}

\usepackage{graphicx}
\usepackage{dcolumn}
\usepackage{bm}
\usepackage{hyperref}
\newcommand{\figurewidth}{\columnwidth}
\begin{document}
\title{Interference effects in the photorecombination of argonlike Sc$^{3+}$ ions:\\
Storage-ring experiment and theory}

 \author{Stefan Schippers}
 \email[e-mail:~]{Stefan.E.Schippers@strz.uni-giessen.de}
 \author{Stefan Kieslich}
 \author{Alfred M\"{u}ller}
 \affiliation{Institut f\"ur Kernphysik, Strahlenzentrum der Justus-Liebig-Universit\"at, 35392
 Giessen,Germany}
 \homepage[homepage:~]{http://www.strz.uni-giessen.de/~k3}

 \author{Gerald Gwinner}
 \author{Michael Schnell}
 \author{Andreas Wolf}
 \affiliation{Max-Planck-Institut f\"ur Kernphysik, 69117 Heidelberg, Germany}
 \homepage[homepage:~]{http://www.mpi-hd.mpg.de/ato/tsrexp.html}

 \author{Aaron Covington}
 \affiliation{Department of Physics, University of Nevada, Reno, NV 89557, USA}

 \author{Mark E. Bannister}
 \affiliation{Physics Division, Oak Ridge National Laboratory, Oak Ridge, TN 37831, USA}

 \author{Li-Bo Zhao}
 \affiliation{ CCAST (World Laboratory), P. O. Box 8730, Beijing 100080, China and\\
 Institute of Theoretical Physics, Chinese Academy of Sciences, Beijing 100088, China}

\date{\today}

\begin{abstract}
Absolute total electron-ion recombination rate coefficients of argonlike Sc$^{3+}$(3s$^2$3p$^6$)
ions have been measured for relative energies between electrons and ions ranging from 0 to 45~eV.
This energy range comprises all dielectronic recombination resonances attached to 3p$\to$3d and
3p$\to$4s excitations.  A broad resonance with an experimental  width of $0.89\pm0.07$~eV due to
the 3p$^5$3d$^2$\,$^2$F intermediate state is found at $12.31\pm0.03$~eV with a small experimental
evidence for an asymmetric line shape. From R-Matrix and perturbative calculations we infer that
the asymmetric line shape may not only be due to quantum mechanical interference between direct and
resonant recombination channels as predicted by Gorczyca \emph{et al.} [Phys. Rev. A {\bf 56}, 4742
(1997)], but may partly also be due to the interaction with an adjacent overlapping DR resonance of
the same symmetry. The overall agreement between theory and experiment is poor. Differences between
our experimental and our theoretical resonance positions are as large as 1.4~eV. This illustrates
the difficulty to accurately describe the structure of an atomic system with an open 3d-shell with
state-of-the-art theoretical methods. Furthermore, we find that a \emph{relativistic} theoretical
treatment of the system under study is mandatory since the existence of experimentally observed
strong 3p$^5$3d$^2$\,$2$D and 3p$^5$3d4s\,$^2$D resonances can only be explained when calculations
beyond LS-coupling are carried out.
\end{abstract}


\maketitle

\section{Introduction\label{sec:intro}}

Dielectronic recombination (DR)
\begin{equation}
\label{eq:dr} e^- + A^{q+} \to [A^{(q-1)+}]^{**} \to  A^{(q-1)+} +  h\nu
\end{equation}
is a two-step electron-ion collision process, where in a first step a multiply-excited intermediate
state is created by a resonant dielectronic capture (DC, inverse Auger) and in a second step that
intermediate state decays by photon emission. Another possible recombination mechanism is radiative
recombination (RR)
\begin{equation}
\label{eq:rr} e^- + A^{q+} \to  A^{(q-1)+} +  h\nu
\end{equation}
which by emission of a photon directly proceeds to a non-autoionizing state of the recombined ion.
When both the initial and the final state including the photons emitted are the same, RR and DR are
indistinguishable and they may interfere, with the signature of interference being an asymmetric DR
resonance line profile in the energy differential recombination cross section
\cite{Alber1984,Badnell1992,Zimmermann1997,Behar2000b}. Asymmetric line shapes have experimentally
so far been observed only in partial recombination cross sections of highly charged uranium ions
\cite{Knapp1995a}. Such an observation is interesting from a fundamental point of view since it
questions the widely used independent processes approximation (IPA) in the calculation of RR+DR
recombination cross sections.

One motivation for investigating the recombination of argonlike Sc$^{3+}$(3s$^2$3p$^6$) ions with
free electrons is the theoretical predication by Gorczyca \emph{et al.}~\cite{Gorczyca1997} of a
strongly asymmetric line shape for the 3p$^5$3d$^{2~2}$F DR resonance caused by interference
between the recombination pathways
\begin{eqnarray}
   \mathrm{e}^- + \mathrm{Sc}^{3+}(3\mathrm{s}^2
3\mathrm{p}^6) &~ \to ~ &
    \mathrm{Sc}^{2+}(3\mathrm{s}^2 3\mathrm{p}^5
3\mathrm{d}^2)\nonumber \\ ~ & \searrow   & \downarrow\label{eq:pathways}\\ ~ &~ &
\mathrm{Sc}^{2+}(3\mathrm{s}^2 3\mathrm{p}^6 3\mathrm{d}) +
       \mathrm{h}\nonumber\nu.
\end{eqnarray}
Theoretically they predicted a Sc$^{2+}$(3s$^2$3p$^5$3d$^{2~2}$F) resonance position of 14.6~eV and
a width of 1.8~eV. The large width of the resonance is explained by the fact that it decays
predominantly via a very fast Super-Coster-Kronig transition (reverse of the horizontal arrow in
Eq.~\ref{eq:pathways}) with a calculated rate of $2.84\times10^{15}$~s$^{-1}$.

A first attempt to experimentally determine the line shape of the Sc$^{2+}$(3p$^5$3d$^{2~2}$F)
resonance by measuring the total Sc$^{3+}$ recombination rate coefficient at a heavy-ion storage
ring failed due to the rather large statistical uncertainty of that measurement
\cite{Schippers1999a} (see also Sec.~\ref{sec:exp}). However, distinct differences between
theoretical and experimental peak positions were found and attributed to the fact that a correct
theoretical description of the highly correlated 3p$^5$3d$^2$ configurations is extremely
difficult. Using Cowan's \cite{Cowan1981} atomic structure code we find that the
Sc$^{2+}$(3p$^5$3d$^2$) configuration gives rise to 45 fine structure components distributed over
an energy interval of about 20~eV width. In order to account for correlation effects Hansen and
Quinet \cite{Hansen1996} considered the interaction of 10 initial and 16 final configurations in
the calculation of 3p$^6$3d\,$\to$\,3p$^5$3d$^2$ transition energies in Ca$^+$. Still their result
for the Ca$^+$(3p$^6$3d$^{~2}$D $\to$ 3p$^5$3d$^{2~2}$F) transition energy deviates 0.76~eV  from
the experimental result (29.34~eV) determined by photoionization of metastable Ca$^+$(3p$^6$3d)
ions \cite{Mueller2001}. This again illustrates the difficulty to accurately describe inner shell
transitions in atomic or ionic systems with open 3d-subshells. In this situation DR measurements
(and complementary photoionization measurements) can yield valuable spectroscopic information on
such systems.

Here we present new experimental as well as theoretical results for the total Sc$^{3+}$
recombination rate coefficient. The experiment is a twofold extension of our previous work
\cite{Schippers1999a}. The new data have much smaller statistical uncertainties and they are
extended to the much wider energy range 0--45~eV (previous range 12.2--18.2~eV) including all DR
series limits due to 3p\,$\to$\,3d and 3p\,$\to$\,4s core excitations.

We have also carried out both nonperturbative and perturbative calculations for the $e^-$ +
Sc$^{3+}$ photorecombination. The nonperturbative method is based on the rigorous continuum-bound
transition theory and the close-coupling R-matrix approach \cite{Zhao2000}. The perturbative
evaluation is a treatment including radiative recombination (RR), dielectronic recombination (DR)
and their interference. Our calculations reveal that three adjacent resonances mask the low energy
side of the 3p$^5$3d$^2$\,$^2$F resonance. This makes the experimental observation of the predicted
\cite{Gorczyca1997} interference between the 3p$^5$3d$^2$($^3$F)\,$^2$F DR resonance and the
continuous RR cross section difficult. Moreover, we find that several experimental DR resonances
not reproduced by the LS-coupling R-matrix calculations of Gorczyca \emph{et al.}
\cite{Gorczyca1997} are due to relativistic effects.

In Sec.~\ref{sec:exp} and Sec.~\ref{sec:theo} we outline our experimental and theoretical
methods. Experimental and theoretical results are presented and discussed in
Secs.~\ref{sec:resexp} and \ref{sec:restheo}, respectively. In Sec.~\ref{sec:3p53d2}
possible origins for the asymmetry of the 3p$^5$3d$^2$($^3$F)\,$^2$F DR resonance are
explored by theoretically considering interference between DR and RR as well as
interference between overlapping DR resonances. Our main results are summarized in
Sec.~\ref{sec:conc}.

\section{Experiment}\label{sec:exp}

The experiment was performed at the heavy ion storage ring TSR of the Max Planck Institut f\"{u}r
Kernphysik (MPIK) in Heidelberg. Details of the experimental setup and the data reduction
procedures can be found in Refs.~\cite{Kilgus1992,Lampert1996,Schippers2000b,Schippers2001c}. The
$^{45}$Sc$^{3+}$ ions were accelerated by the MPIK tandem booster facility to their final energy
and subsequently injected into the TSR. The rather low charge to mass ratio of $q/A=1/15$ leads to
the rather unfavorable condition (see below) that the ion energy is limited by the maximum bending
power of the TSR dipole magnets. With the maximum magnetic rigidity of $B\rho = 1.4$~Tm the highest
accessible ion energy can be calculated from
\begin{equation}
E_\mathrm{i}/A \approx 931.5\,\textrm{MeV/u}\,
\left[\sqrt{1\!+\!\left(\frac{q}{A}\frac{B\rho}{3.107\,\textrm{Tm}} \right)^2}-1
\right]\label{eq:Eion}
\end{equation}
to $E_\mathrm{i}/A \approx 420$~keV/u. In the storage ring the circulating 18.9~MeV Sc$^{3+}$ ion
beam was merged with the magnetically guided electron beam of the electron cooler. During
electron-cooling, the electrons have to move with the same average velocity as the ions. This
condition defines the cooling energy $E_\mathrm{c} = (E_\mathrm{i}/A) (m_\mathrm{e}/\mathrm{u})
\approx 230$~eV. In order to increase the phase space density of the already stored ions and
thereby free phase space for the next injection pulse from the accelerator, electron-cooling was
allowed to be effective for 2 seconds after each injection of ions. These `ecool-stacking' cycles
\cite{Grieser1991} were repeated three times until the accumulated ion current stabilized at values
of up to 5~$\mu$A. Before data taking was started a prolonged cooling interval of 5~s after the
last injection allowed the ion beam to shrink to a final diameter of about 2~mm as verified with a
beam profile monitor based on residual gas ionization \cite{Hochadel1994a}.

During the measurement the electron energy $E_\mathrm{e}$ in the laboratory frame was stepped
through a preset range of values different from $E_\mathrm{c}$ thus introducing non-zero relative
energies
\begin{equation}
E_\mathrm{rel} = \left(\!\sqrt{E_\mathrm{e}}-\sqrt{E_\mathrm{c}}\,\right)^2\label{eq:erel}
\end{equation}
between ions and electrons. In the data analysis a relativistically correct expression
\cite{Schippers2000b} was used instead of Eq.~\ref{eq:erel}. Recombined Sc$^{2+}$ ions were counted
with $100{+0 \atop -3}$\% efficiency as a function of cooler voltage on a single particle detector
\cite{Rinn1982} located behind the first dipole magnet downstream of the electron cooler. The
dipole magnet bends the circulating Sc$^{3+}$ ion beam onto a closed orbit and separates the
recombined Sc$^{2+}$ ions from that orbit. In between two measurement steps the cooler voltage was
first set back to the cooling value in order to maintain the ion beam quality and then set to a
reference value at $E_\mathrm{rel} = 45$~eV which is chosen to lead to a relative velocity where
the electron-ion recombination coefficient is only due to a negligible RR rate. Under this
condition the recombination rate measured at the reference point monitors the background signal.
Choosing short time intervals of 10~ms duration for dwelling on the measurement, cooling and
reference voltages ensured that the experimental environment did not change significantly in
between signal and background measurement. An additional interval of 1.5~ms after each change of
the cooler voltage allowed the power supplies to reach the preset values before data taking was
started.

At the rather low ion energy of 420~keV/u the cross section for electron capture from residual gas
molecules is estimated --- taking the measured residual gas composition into account and using a
semi-empirical formula \cite{Schlachter1983} for the charge capture cross section
$\sigma^\mathrm{(capt)}$ --- to be as large as $\sigma^\mathrm{(capt)} = 4\times10^{-18}$~cm$^2$.
Electron capture from residual gas molecules is therefore expected to contribute significantly to
the measured recombination signal as a background even at the TSR residual gas pressure of only
$5\times10^{-11}$~mbar. This already proved to be all the more the case in a previous TSR
experiment with Sc$^{3+}$ ions \cite{Schippers1999a} where due to a technical defect in one of the
TSR dipole power supplies a reduced maximum rigidity of only $B\rho \approx 1.2$~Tm had  been
available. Consequently, the ion energy had been limited to 300~keV/u. Because of the rapid
increase of $\sigma^\mathrm{(capt)}$ with decreasing ion energy and because the experiment had been
carried out in early summer, i.\,e., close to a maximum in the slight seasonal variation of the TSR
residual gas pressure, the measured signal to background ratio had been of the order of 1/100.
Under these conditions the Sc$^{3+}$ recombination rate coefficient could only be measured over a
limited energy range with large statistical uncertainties. In the present measurements, that were
carried out in December 2000, the signal to background ratio was a factor of about 10 higher as
compared to the previous experiment.

The electron-ion recombination coefficient $\alpha (E_\mathrm{rel})$ is obtained as the
background-subtracted recombination count rate normalized to the electron current and to the number
of stored ions \cite{Kilgus1992}. The estimated systematical uncertainties are 15\% for the
absolute value of the measured rate coefficient and less than 2\% for the relative energy in the
energy range under study.

\section{Theory}\label{sec:theo}

\subsection{Nonperturbative approach of continuum-bound transitions}

Starting from the rigorous continuum-bound transition theory,
Davies and Seaton \cite{Davies1969} discussed the process of
emission of radiation into the optical continuum due to radiative
capture of an electron by an ion and derived a general formalism
including radiation damping. The details of their formalism and
our numerical scheme for application can be found in
Ref.~\cite{Davies1969} and Ref.~\cite{Zhao2000}, respectively.
Nevertheless, a brief review of the formalism and our method
follows.

Using the close-coupling R-matrix approach (see \cite{Zhao2000} for details) we start from a
treatment of the electron-ion collision process that neglects any interaction with the radiation
field. The resulting wave functions are used as a basis for setting up equations including the
interaction with the radiation field in the approximation that only the electric-dipole terms are
retained and that the radiation field is restricted to one-photon and no-photon states. The exact
solutions for the probability amplitudes of the time-dependent matrix equations involved can be
obtained by the application of a Laplace transform. They are expressed in terms of a scattering
matrix $\chi$ with the partitioning
\begin{equation}\label{Smatrix}
\chi=\left(\begin{array}{cc}
{\chi}_\mathrm{ee} & {\chi}_\mathrm{ep} \\
{\chi}_\mathrm{pe} & {\chi}_\mathrm{pp}
\end{array}
\right)
\end{equation}
with ${\chi}_\mathrm{ee}$ representing the submatrix for electron-electron scattering allowing for
radiative decays, ${\chi}_\mathrm{ep}$ that for photoionization, ${\chi}_\mathrm{pe}$ that for
electron capture with the emission of a photon, and ${\chi}_\mathrm{pp}$ that for photon-photon
scattering. ${\chi}_\mathrm{ee}$ and ${\chi}_\mathrm{pe}$ are written as
\begin{eqnarray}
{\chi}_\mathrm{ee}&=&{ S} [1-2\pi^2{ D}(1+{ Z})^{-1}{{ D}^{\dagger}}],\label{See}\\
{\chi}_\mathrm{pe}&=&-2\pi \mathit{i}(1+{ Z})^{-1}{ D}^{\dagger},\label{Spe}
\end{eqnarray}
respectively, where ${ S}$ is the usual electron-electron scattering matrix in the absence of
interaction with radiation fields, and ${D}$ is the reduced dipole matrix (${D}^\dagger$ denotes
its hermitian conjugate) with its matrix element in the form
\begin{equation}\label{Dmatrix}
{D}_{{\gamma}J,{\gamma}^{\prime}J^{\prime}}= \left(\frac{2\omega ^3\alpha^3}{3\pi} \right)^{1/2}
\frac{\langle\gamma J\parallel R\parallel\gamma^{\prime} J^{\prime}\rangle}{(2J+1)^{1/2}}
\end{equation}
where $\alpha$ is the fine-structure constant, $\omega$ is the photon energy in units of hartrees,
and $R=\sum_i {\bf r}_i$ is the dipole operator with the summation extending over all atomic
electrons. The quantum numbers ${\gamma}J$ and ${\gamma}^{\prime}J^{\prime}$, respectively, specify
the continuum and bound states of the atomic system, and $J(J^\prime)$ are the total angular
momenta. The wave function of the continuum electron is normalized per unit hartree.  In
Eqs.~\ref{See} and \ref{Spe} the matrix
\begin{equation}\label{Zmatrix}
{Z}(\Omega)= - \mathit{i}\pi\!\!\int dE \frac {D^{\dagger}(E) D(E)}{(E-\Omega)}
\end{equation}
is related to radiation damping. The variable $\Omega$ denotes the total energy of the ion+photon
system. In the usual first-order theory damping is neglected and $\chi_\mathrm{pe}=
-2\pi\mathit{i}{D}$. For the calculation of the PR cross section with damping we employ the
numerical method developed by Zhao et al.\ \cite{Zhao2000} for the evaluation of the principal
value of the integral appearing in Eq.~\ref{Zmatrix}.

In the present calculation the parameters for the 1s, 2s, 2p, 3s, 3p orbitals of scandium were
taken from the compilation of Clementi and Roetti \cite{Clementi1974}, and the 3d orbital was
optimized on the 3p$^5$3d\,$^{1,3}$P$^\mathrm{o}$, $^{1,3}$D$^\mathrm{o}$, $^{1,3}$F$^\mathrm{o}$
states (weighted) by using the CIV3 code of Hibbert \cite{Hibbert1975}. We also evaluate PR cross
sections for further target states including 4s, 4p, 4d orbitals. All these orbitals were optimized
in a way similar to the 3d orbital. The symmetries $^{2S+1}L^{\pi}$ with $L\leq 5$ were included in
our calculations. Here, S and L are the total spin and orbital angular momentum quantum numbers,
respectively, and $\pi$ denotes the parity. In our calculation we included configurations of odd
parity only, since these are expected to lead to the dominant resonances in the energy region of
our primary interest around the 3p$^5$3d$^2$($^3$F)\,$^2$F resonance.

\subsection{Perturbation theory}\label{sec:pert}

On the basis of the principle of detailed balance, the photorecombination (PR) cross sections (in
atomic units) from an initial continuum state $j$ to a final bound state $f$ may be written in the
form \cite{Sobelman}
\begin{equation}\label{Sigpr}
\sigma^\mathrm{PR}_{jf}=\frac{g_f}{g_j}\frac{\alpha^2\omega^2}{2\epsilon} \sigma^\mathrm{PI}_{fj}
\end{equation}
where $g_j$ and $g_f$ are the statistical weights of the initial ion core and the final recombined
ion, $\omega$ and $\epsilon$ are the photon energy and the free electron energy in hartrees, and
$\sigma^\mathrm{PI}_{fj}$ denotes the photoionization (PI) cross section from state $f$ to state
$j$. In first order perturbation treatment it is
\begin{equation}\label{Sigpi}
\sigma^\mathrm{PI}_{fj}=\frac{4\pi^2\alpha\omega}{3}|M_{fj}|^2
\end{equation}
where $M_{fj}$ is the PI matrix element of the corresponding transition with the continuum state
being normalized per unit energy (hartree). According to continuum-bound configuration-interaction
theory \cite{Fano1961}, the perturbative PI matrix element $M_{fj}$ in low orders may be written in
the form (see also \cite{Gorczyca1997,Mitnik1999}),
\begin{eqnarray}\label{Melement}
M_{fj}&=&\langle j|R|f\rangle\left(1-i\sum_d\frac{\Gamma^\mathrm{a}_{dj}/2}{\Delta_d+i\Gamma_d/2}\right)  \nonumber\\
&& +\sum_d\frac{\langle j|V|d\rangle\langle d|R|f\rangle}{\Delta_d+i\Gamma_d/2}
\end{eqnarray}
where $\Gamma_d$ is the summation of radiative and autoionization widths of the resonance state
$d$, $\Gamma^\mathrm{a}_{dj}$ is the autoionization width from state $d$ to state $j$,
$\Delta_d=\epsilon - \epsilon_d$, in which $\epsilon_d$ is the energy level of resonance $d$, $V$
is the electron-electron interaction. It should be noted that in the usual independent-processes
and isolated-resonance approximations all cross terms in $|M_{fj}|^2$ are omitted that lead to
interference between resonant and non-resonant recombination channels as well as to the interaction
between resonances. In the present investigation these cross terms are taken into account where
necessary.

\section{Results and Discussion}\label{sec:res}

\subsection{Experimental results}\label{sec:resexp}

\begin{figure}[bbb]
\begin{center}
\includegraphics[width=\figurewidth]{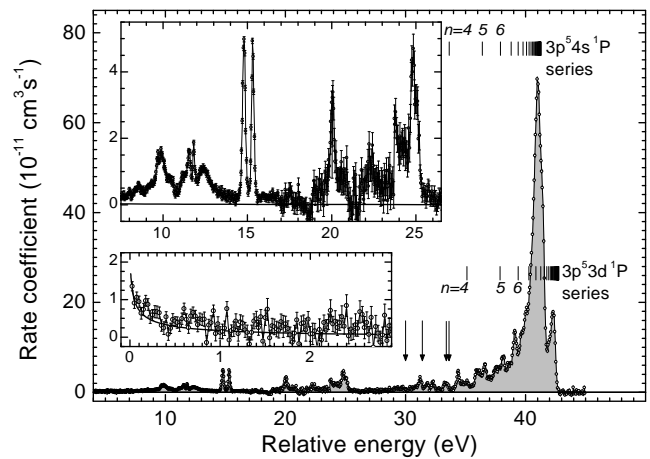}
\end{center}
\caption{\label{fig:expall}Measured Sc$^{3+}$(3s$^2$3p$^6$) recombination-rate coefficient.
Vertical lines denote resonance positions of the 3p$^5$4s$^{~1}$P\,$nl$ and 3p$^5$3d$^{~1}$P\,$nl$
Rydberg series of DR resonances. Vertical arrows denote possible positions of further 3p$^5$3d
series limits (see Tab.~\ref{tab:slim}). The insets enlarge the spectrum at low energies. Different
energy intervals were measured for different time durations and therefore exhibit different
statistical uncertainties.  The rise of the recombination-rate coefficient towards $E_\mathrm{rel}
= 0$~eV is due to RR. The full line approaching $\alpha(E_\mathrm{rel}) = 0$ towards higher
energies is the total hydrogenic RR rate coefficient (see text).}
\end{figure}

\begin{table}[bbb]
 \caption{\label{tab:slim}3p-excitation energies of argonlike Sc$^{3+}$(3s$^2$3p$^6$) taken from the NIST atomic
 spectra database \protect\cite{Martin1999}. Note that the excitation to the 3p$^5$4s terms
 requires less energy than the excitation to the highest 3p$^5$3d term.}
 \begin{ruledtabular}
 \begin{tabular}{rrrd}
 Configuration & Term & $J$ & \multicolumn{1}{c}{Energy (eV)}\\
 \hline
 3s$^2$3p$^5$3d     & $^3$P   &  0  & 29.72190\\
                    &         &  1  & 29.80618\\
                    &         &  2  & 29.98113\\
 3s$^2$3p$^5$3d     & $^3$F   &  4  & 31.08387\\
                    &         &  3  & 31.25746\\
                    &         &  2  & 31.41831\\
 3s$^2$3p$^5$3d     & $^1$D   &  2  & 33.15641\\
 3s$^2$3p$^5$3d     & $^3$D   &  3  & 33.23202\\
                    &         &  1  & 33.40464\\
                    &         &  2  & 33.40878\\
 3s$^2$3p$^5$3d     & $^1$F   &  3  & 33.60660\\
 3s$^2$3p$^5$4s     & $^3$P   &  2  & 41.29801\\
                    &         &  1  & 41.46096\\
                    &         &  0  & 41.82486\\
 3s$^2$3p$^5$4s     & $^1$P   &  1  & 41.84264\\
 3s$^2$3p$^5$3d     & $^1$P   &  1  & 42.77523\\
 \end{tabular}
 \end{ruledtabular}
\end{table}

Fig.~\ref{fig:expall} shows the measured Sc$^{3+}$(3s$^2$3p$^{6~1}$S) recombination rate
coefficient. The most prominent feature is the peaklike structure around $E_\mathrm{rel}=40$~eV. It
can be attributed to mainly unresolved high $n$ Rydberg DR resonances attached to 3p$^5$3d$^{~1}$P
and 3p$^5$4s$^{~1, 3}$P core excitations. Vertical lines indicate DR resonance positions $E(n)$ as
calculated with the Rydberg formula $E(n) = E(\infty) - {\cal R}\,q^2/n^2$ with ${\cal R} =
13.606$~eV, $q=3$, and the series limits $E(\infty)$ taken from Tab.~\ref{tab:slim}.  It should be
noted that we do not observe Rydberg resonances to arbitrary high $n$, since recombined
Sc$^{2+}$(3p$^6$\,$nl$) ions in weakly bound high $n$ states are field-ionized by motional electric
fields in the charge-analyzing dipole magnet and therefore do not reach the recombination detector.
Applying a detailed model of the field-ionizing properties of our apparatus that takes into account
the radiative decay of high Rydberg states on their way from the electron cooler to the
charge-analyzing magnet as well as state-selective field-ionization probabilities
\cite{Schippers2001c} we estimate the highest detected Rydberg state to be approximately $n=20$.

In principle, Rydberg series attached to further 3p$^5$3d core excitations (Tab.~\ref{tab:slim})
could have been expected to be visible. The respective core-excitation energies are marked by
vertical arrows in Fig.~\ref{fig:expall}. However, no strong DR resonances belonging to such series
are observed. Apparently, only dipole allowed (under $LS$-coupling conditions) $^1$S\,$\to$\,$^1$P
core excitations give rise to strong DR channels. This has already been observed for iso-electronic
Ti$^{4+}$ ions \cite{Schippers1998} where DR proceeds almost exclusively via resonances attached to
3p$^5$3d$^{~1}$P excitations. In contrast to the present finding for Sc$^{3+}$, in the DR spectrum
of Ti$^{4+}$ no DR resonances due to 3p$\to$4s excitations were found. One strong hint to the
origin of this difference is the fact that in Sc$^{3+}$ the 3p\,$\to$\,4s excitation energies are
about 1~eV lower than the 3p$^5$3d$^{~1}$P excitation energy (Tab.~\ref{tab:slim}), while in
Ti$^{4+}$ they are higher by up to 6~eV. Although being iso-electronic Sc$^{3+}$ and Ti$^{4+}$ do
have a markedly different structure.

The recombination rate coefficient rises sharply towards zero relative energy (lower inset of
Fig.~\ref{fig:expall}). This rise is due to RR. The RR rate coefficient is rather small as can be
expected from the fact that  RR into the K, L, M$_1$ and M$_{2,3}$ shells of argonlike Sc$^{3+}$ is
not possible. This is in contrast to e.\,g.\ bare Li$^{3+}$ ions where these channels yield the
major contribution (43\% at $E_\mathrm{rel} = 10^{-6}$~eV) to the total hydrogenic RR cross section
\cite{Burgess1964b} summed up to $n=20$. The hydrogenic total Sc$^{3+}$ rate coefficient, i.\,e.\
the sum of the $nl$-differential Li$^{3+}$ RR cross sections ranging from the 3d-subshell to the
experimental cutoff quantum number $n=20$ convoluted with the experimental electron-energy
distribution (see e.\,g.\ Ref.~\cite{Schippers2001c}), is plotted as the full line in the lower
inset of Fig.~\ref{fig:expall}. It agrees with the experimental curve within the experimental
uncertainties. In spite of the long measuring time of 60 hours, in particular in the energy range
of 8--17~eV, the statistical uncertainty did not become low enough to bring out all structures. At
the present level of statistical uncertainty it cannot be decided whether the faint structures in
the energy range 1--7.5~eV are due to weak DR resonances or due to unaccounted background
modulations.

The DR resonances visible in the upper inset of Fig.~\ref{fig:expall} are due to 3p$^5$3d$^2$ and
3p$^5$3d\,4$l$ doubly excited configurations. These configurations are considerably shifted down in
energy as compared to the Rydberg energies shown by vertical lines in Fig.~\ref{fig:expall} and the
energies of the various terms must be taken from detailed calculations. As already mentioned in the
introduction the terms belonging to the 3p$^5$3d$^2$ configuration straddle an energy range of
about 20~eV. Also the 3p$^5$3d\,4$l$ configurations can be expected to be subject to strong fine
structure interactions and hence to be spread out over several electron volts. The energy range
between  about 8 and 26~eV (upper inset in Fig.~\ref{fig:expall}) was searched preferentially in
order to detect the broad 3p$^5$3d$^2$\,$^2$F Super-Coster-Kronig resonance predicted by theory.
The measurement reproduces the strong double peak structure near 15~eV found already in the
previous experiment \cite{Schippers1999a} (scan range 12--18~eV) but now also reveals considerable
additional structure both above and below these peaks. While much of this structure seems to arise
from a superposition of several narrow peaks, a single broad resonance (with some narrower
structure superimposed on its low energy slope) seems to dominate the recombination rate near
12.3~eV. This signal bears strong similarity with the predicted 3p$^5$3d$^2$\,$^2$F
Super-Coster-Kronig resonance \cite{Gorczyca1997} and we take this feature as a key to assigning a
considerable part of the observed spectrum. This interpretation is also supported by the argument
that for the isoelectronic system Ti$^{4+}$ \cite{Schippers1998} the 3p$^5$3d$^2$\,$^2$F resonance
was observed about 3~eV below the prediction obtained by the same theoretical approach
\cite{Gorczyca1998a}, so here a position of $\sim 12$~eV can be expected.

\begin{figure}[ttt]
\begin{center}
\includegraphics[width=\figurewidth]{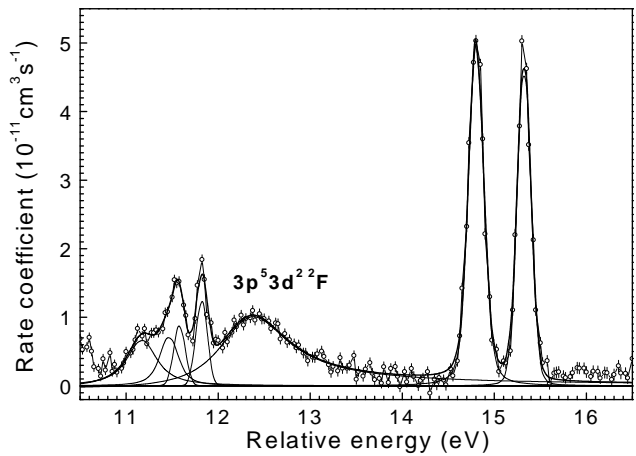}
\end{center}
\caption{\label{fig:expfit} Fit of Lorentzian and Fano line profiles to the experimental data over
the energy range 10.9--16.5~eV. The individual line profiles have been convoluted with a Gaussian
representing the experimental energy distribution (see appendix). The fit results are summarized in
Tab.~\protect\ref{tab:fit}. The fit delivers a significant asymmetry ($Q=6.3\pm1.8$) of the broad
resonance at 12.3~eV. Moreover, from the fit a longitudinal electron temperature
$k_\mathrm{B}T_{\|} = 0.15\pm0.03$~meV is obtained.}
\end{figure}

\begin{table}[bbb]
 \caption{\label{tab:fit}Results obtained from fitting Lorentzian (Eq.~\ref{eq:lor}) and
 Fano (Eq.~\ref{eq:Fano}) line profiles (convoluted with a Gaussian, see appendix) to resonances in the energy range
 10.9--16.5~eV (see Fig.~\protect\ref{fig:expfit}). The errors given do not include systematic uncertainties
 in the energy calibration ($\pm 2\%$) and the absolute rate coefficient determination ($\pm 15\%$).}
 \begin{ruledtabular}
 \begin{tabular}{r@{${}\pm{}$}lr@{${}\pm{}$}ld@{${}\pm{}$}l}
 \multicolumn{2}{c}{$E_d$} & \multicolumn{2}{c}{$\Gamma$} & \multicolumn{2}{c}{$A$} \\
 \multicolumn{2}{c}{(eV)} & \multicolumn{2}{c}{(eV)} & \multicolumn{2}{c}{(10$^{-12}$cm$^3$s$^{-1}$eV)} \\
 \hline
 11.175  & 0.041 & 0.317 & 0.099 & 3.8 &  1.8 \\
 11.481  & 0.143 & 0.140 & 0.253 & 2.6 &  6.1 \\
 11.587  & 0.036 & 0.000 & 0.193 & 1.2 &  3.8 \\
 11.819  & 0.004 & 0.000 & 0.021 & 1.9 &  0.4 \\
 12.305  & 0.027 & 0.889 & 0.066 & 14.3&  1.0 \footnote{Fano profile (Eq.~\protect\ref{eq:Fano}) with $Q=6.3 \pm 1.8$.} \\
 14.802  & 0.001 & 0.056 & 0.006 & 11.6&  0.2  \\
 15.324  & 0.001 & 0.018 & 0.006 & 9.1 &  0.2  \\
 \end{tabular}
 \end{ruledtabular}
\end{table}

A closeup of the experimental spectrum is shown in Fig.~\ref{fig:expfit}, where the resonances in
the energy range 10.9--16.5~eV have been fitted by either Lorentzian line profiles
\begin{equation}\label{eq:lor}
L(E) = \frac{A}{\Gamma_d}\frac{2}{\pi}\frac{1}{1+\varepsilon^2}
\end{equation}
or an asymmetric Fano profile
\begin{equation}\label{eq:Fano}
F(E) = \frac{A}{Q^2\Gamma_d}\frac{2}{\pi}\left[\frac{(Q+\varepsilon)^2}{1+\varepsilon^2}-1\right]
\end{equation}
with $\varepsilon = 2(E\!-\!E_d)/\Gamma_d$. The term -1 in the square brackets ensures that
$F(E)\to 0$ for $E\to\pm\infty$ and the normalization factors have been chosen such that $F(E)\to
L(E)$ in the limit $Q\to \infty$. Conversely, small absolute values of the asymmetry parameter $Q$
lead to significantly asymmetric line shapes. Assuming that the dominant contribution to the
resonance width $\Gamma_d$ stems from autoionization and neglecting contributions from neighboring
resonances, $Q$ can be expressed in terms of the matrix elements occurring in Eq.~\ref{Melement},
i.\,e.,
\begin{equation}\label{eq:Q}
Q = \frac{2\,\langle j\vert V\vert d\rangle\,\langle d\vert  R\vert f\rangle}{\Gamma_d\, \langle
j\vert  R\vert f\rangle}
\end{equation}
as can be shown by simple algebraic manipulations on $\vert M_{fj}\vert^2$. For each resonance the
parameters varied in the fit were the peak area $A$, the resonance energy $E_d$, the resonance
width $\Gamma_d$, and in case of the 3p$^5$3d$^{2~2}$F resonance the asymmetry parameter $Q$. In
the fit both $L(E)$ and $F(E)$ have been convoluted (see appendix) by a Gaussian representing the
experimental energy-distribution function with its full width at half maximum (FWHM) $\Delta E
\approx 4(E_d \, k_\mathrm{B} T_{\|} \ln 2)^{1/2} $ in the energy region of interest
\cite{Kilgus1992}. The longitudinal electron-beam temperature $k_\mathrm{B}T_{\|}$ ($k_\mathrm{B}$
denotes the Boltzmann constant) has also been allowed to vary during the fit. From the fit we
obtain $k_\mathrm{B}T_\| = 0.16\pm0.03$~meV. Further fit results are summarized in
Tab.~\ref{tab:fit}.

The most important result of the fit is the determination of the 3p$^5$3d$^{2~2}$F resonance energy
and width to $12.31\pm 0.03$~eV ($\pm 2\%$ systematic uncertainty) and $0.89\pm 0.07$~eV,
respectively. For the asymmetry parameter of this resonance we obtain $Q=6.3\pm1.8$. A major source
of uncertainty beyond the errors given is the background from collisions with residual gas
molecules. Under the conditions of the present experiment the background causes the major fraction
of the measured recombination count rate (see Sec.~\ref{sec:exp}). A subtraction of this high
background level from the measured signal is difficult, especially, if changes of the residual gas
pressure have occurred on the time scale of the switching between signal and background
measurement. In this case a small fraction of the background may not have been properly subtracted
possibly leading, e.\,g., to the nonzero recombination signal in the energy range 15.6--17~eV. In
principle, this signal might also be caused by a group of weak DR-resonances, however, this
assumption is not supported by our theoretical calculations (see below). When allowing a constant
background level as an additional free parameter in the fit a considerably different asymmetry
parameter $Q=10.4\pm4.1$ is obtained with the fitted background level amounting to $6.0\pm 0.9
\times 10^{-13}$~cm$^3$s$^{-1}$. On the other hand the resonance position and width change only
within their errors as listed in Tab.~\ref{tab:fit} to $12.35\pm 0.03$~eV and $0.85\pm 0.07$,
respectively.

The significance of our fit result for the asymmetry parameter may also be assessed by fitting the
3p$^5$3d$^{2~2}$F resonance with a \emph{symmetric} Lorentzian instead of an \emph{asymmetric} Fano
line shape. From such a Lorentzian fit we obtain a $\chi^2$ value that is higher by only 5\% as
compared to the Fano fit, i.\,e., the experimental evidence for an asymmetric line shape is not
very strong.

\begin{figure}[ttt]
\begin{center}
\includegraphics[width=\figurewidth]{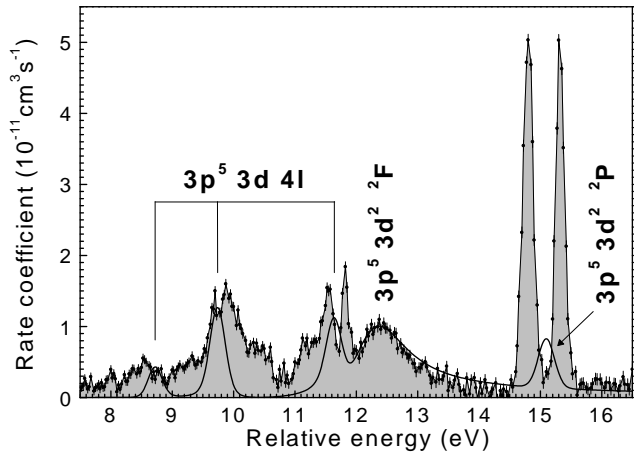}
\end{center}
\caption{\label{fig:expgor} Comparison of the measured Sc$^{3+}$ rate coefficient with the
theoretical result of Gorczyca \emph{et al.}~\protect\cite{Gorczyca1997} (full curve) in the region
of the 3p$^5$3d$^{2~2}$F resonance. The theoretical peak designations are given. The theoretical
rate coefficient has been shifted by 2.2~eV to lower energies and multiplied by 0.63 in order to
obtain agreement with the measured 3p$^5$3d$^{2~2}$F resonance position and height.}
\end{figure}

Fig.~\ref{fig:expgor} shows the comparison of the present experimental result with the theoretical
result of Gorczyca \emph{et al.}\ \cite{Gorczyca1997} in the region of the 3p$^5$3d$^{2~2}$F
resonance. As already inferred from our previous work \cite{Schippers1999a} the theoretical
resonance positions do not match the experimental ones. For the comparison presented in
Fig.~\ref{fig:expgor} the theoretical energy scale has been shifted by 2.2~eV in order to match the
experimental position of the broad 3p$^5$3d$^{2~2}$F resonance. Moreover, the absolute theoretical
rate coefficient has been normalized to the experimental height of this resonance by a
multiplication factor of 0.63. After these adjustments a basic similarity between predicted and
measured spectrum is found. In particular, a broad feature is observed strongly resembling the
calculated  3p$^5$3d$^2$\,$^2$F resonance.  However, in the experimental spectrum considerably more
resonances are visible than in the theoretical one. The theoretically predicted single peaks at
11.6 and 15.1~eV appear to be split into two components. Also, the predicted relative resonance
strengths deviate from the experimental findings. The theoretical 3p$^5$3d$^{2~2}$P peak at 15.1~eV
is much smaller than the experimental doublet.

\subsection{Theoretical results}\label{sec:restheo}

\begin{table}[bbb]
\caption{\label{tab:energies}Comparison of experimental and theoretical energy levels of some
doubly excited Sc$^{2+}$ states relative to the 3p$^{6~1}$S state in units of eV. The assignment of
the experimental resonances below 12~eV is uncertain.}
 \begin{ruledtabular}
 \begin{tabular}{ldcccc}
 State & \multicolumn{1}{c}{Exp.} &  \multicolumn{4}{c}{Theory\footnote{A: this work,
 B: Ref.~\protect\cite{Gorczyca1997}, C: Ref.~\protect\cite{Tiwary1983}, D: Ref.~\protect\cite{Altun1999}}} \\
 \cline{3-6} &  &    A & B & C & D \\ \hline
 3p$^5$3d($^3$F)4s\,$^2$F  \footnote{1s$^{2}$2s$^{2}$2p$^{6}$3s$^{2}$ is omitted.}
                                            & 11.18 &  11.00 &       &11.59&  \\
 3p$^5$3d($^1$D)4s\,$^2$D                   & 11.48 &        &       &12.18&  \\
 3p$^5$3d($^1$F)4s\,$^2$F                   & 11.58 &  12.71 &       &12.77&  \\
 3p$^5$3d($^3$D)4s\,$^2$D                   & 11.82 &        &       &12.89&  \\
 3p$^5$3d$^2$($^3$F)\,$^2$F                 & 12.31 &  13.61 & 14.62 &13.23& 15.74\\
 3p$^5$3d$^2$($^3$P)\,$^2$P                 & 14.80 &  16.40 & 17.35 &15.85& 16.81\\
 3$p^5$3d$^2$($^3$F)\,$^2$D                 & 15.32 &        &       &16.90&  \\
\end{tabular}
\end{ruledtabular}
\end{table}

\begin{figure}[ttt]
\begin{center}
\includegraphics[width=\figurewidth]{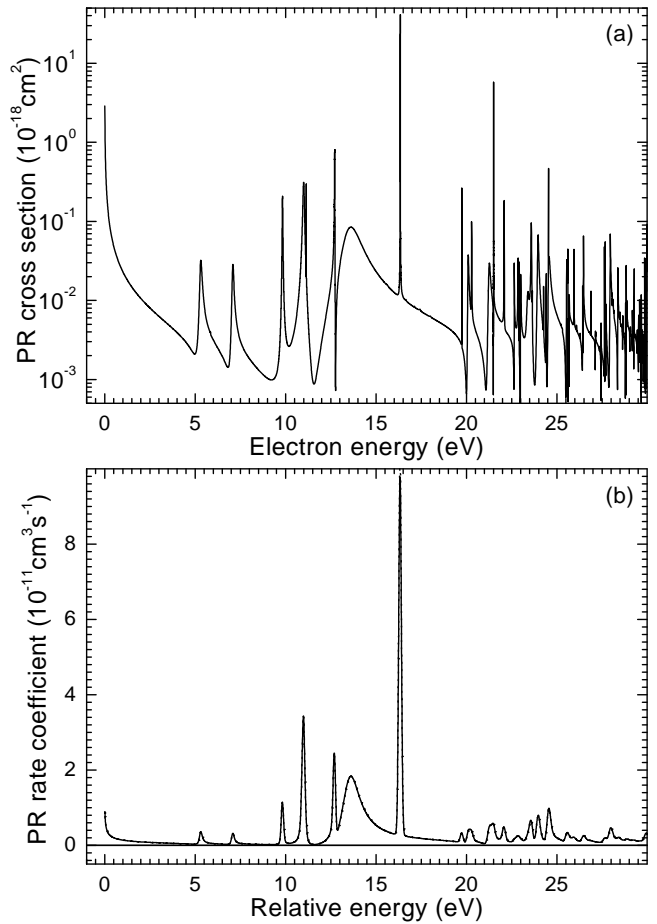}
\end{center}
\caption{\label{fig:csalp}(a) Total Sc$^{3+}$ photorecombination cross section as calculated with
the R-matrix approach in the LS coupling scheme including 3p$^5$3d\,$nl$ DR channels with $n\leq 4$
and with the symmetries $^2L^{\pi}$ with $ L \leq 5$. (b) Merged beams recombination-rate
coefficient derived from the above R-Matrix cross section by convolution with the experimental
electron-energy distribution function with $k_\mathrm{B}T_\|=0.15$~meV and $k_\mathrm{B}T_\perp =
10$~meV.}
\end{figure}

The result of our LS-coupling--R-Matrix calculation of the total Sc$^{3+}$ PR cross section
including damping is plotted in Fig.~\ref{fig:csalp}a. It can be noticed that there is a very wide
resonance at about 13.6 eV. As already predicted by Gorczyca \emph{et al.} \cite{Gorczyca1997}, it
is due to DR via the 3p$^5$3d$^2$($^3$F)\,$^2$F state. However, our resonance energy is by about 1
eV closer to the experimental value of 12.29~eV (see Tab.~\ref{tab:fit}) than that of Gorczyca
\emph{et al}. This difference originates from the use of different sets of basis functions in both
calculations. Table~\ref{tab:energies} lists measured and theoretical resonance energies including
further theoretical results \cite{Tiwary1983,Altun1999}. The relatively large differences between
the various theoretical results illustrates again the difficulty to exactly describe the
many-electron atomic system under consideration.

\begin{figure}[ttt]
\begin{center}
\includegraphics[width=\figurewidth]{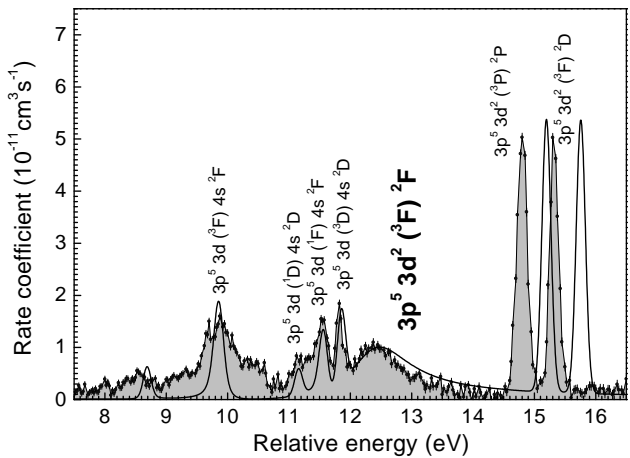}
\end{center}
\caption{\label{fig:expzhao} Comparison of the measured Sc$^{3+}$ rate coefficient with the present
theoretical result (R-Matrix + perturbation theory) in the region of the 3p$^5$3d$^2$($^3$F)\,$^2$F
resonance. The theoretical peak designations are given. The theoretical rate coefficient has been
shifted by 1.13~eV to lower energies and multiplied by 0.55 in order to obtain agreement with the
measured 3p$^5$3d$^2$($^3$F)\,$^2$F resonance position and height.}
\end{figure}

In order to be able to more accurately  compare the theoretical result with the measured
merged-beams rate coefficient we convoluted the R-matrix cross section with the experimental
electron energy distribution function, that can be parameterized by the longitudinal and transverse
(with respect to the electron beam direction) electron beam temperatures $T_\|$ and $T_\perp$,
respectively \cite{Kilgus1992}. For the convolution result displayed in Fig.~\ref{fig:csalp}b we
have used $k_\mathrm{B}T_\| = 0.15$~meV (cf., Fig.~\ref{fig:expfit}) and $k_\mathrm{B}T_\perp =
10$~meV. Comparing with the experimental data in Fig.~\ref{fig:expall} we find that the R-matrix
calculation does not reproduce the experiment in every detail. For example, the calculated
resonance peaks in the energy range 7--17~eV are too high, whereas those beyond 19~eV are too low.
Moreover, some of the observed peaks at lower energies are not reproduced by the R-matrix
calculation. The disagreement at high energies can be traced back to the neglect of even-parity
configurations and to the neglect of relativistic effects by restricting the R-Matrix calculation
to $LS$-coupling. This also explains the missing of peaks at lower energies. Using Cowan's code
\cite{Cowan1981} for the calculation of DR resonances in the framework of perturbation theory
(Sec.~\ref{sec:pert}) we find that due to relativistic effects (mainly spin-orbit interaction) also
$LS$-forbidden excitations to 3p$^5$3d$^2$($^3$F)\,$^2$D as well as to 3p$^5$3d($^1$D)4s\,$^2$D and
3p$^5$3d($^3$D)4s\,$^2$D doubly excited states may contribute to the PR cross section. It should be
emphasized that resonances arising from relativistic effects have in general small Auger widths if
the atomic number $Z$ is not large. As has been explained in some detail by Zhao and Shirai
\cite{Zhao2001} they, nevertheless, may still play an important role in recombination if their
radiative lifetime is of the order of the radiative lifetimes of the dominant DR resonances.

\begin{figure}[ttt]
\begin{center}
 \includegraphics[width=\figurewidth]{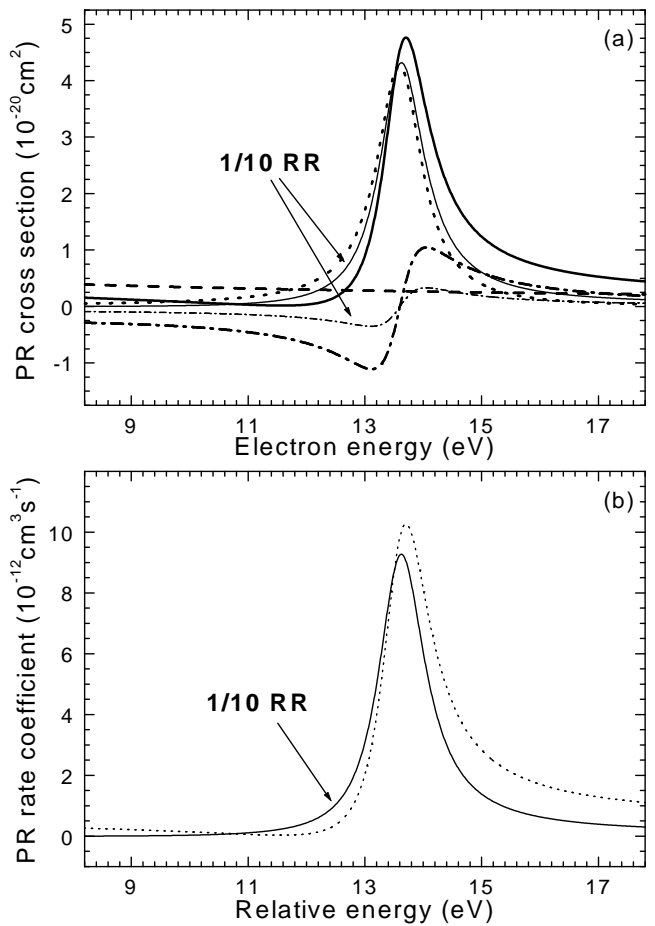}
\end{center}
\caption{\label{fig:drrr}(a) Cross section for recombination of Sc$^{3+}$ via the intermediate
3p$^5$3d$^2$($^3$F)\,$^2$F state into the 3p$^6$3d\,$^2$D state: total cross section (solid line),
nonresonant RR cross section (dashed line), DR cross sections (dotted line), and interference term
(dashed-dotted line). Also given are the total cross section and the interference term for the case
of an artificially reduced (by a factor 10) RR cross section (thinner lines). (b) Merged-beams
recombination rate coefficients corresponding to the calculated (dotted line) and artificial (solid
line) total cross sections.}
\end{figure}

As shown in Fig.~\ref{fig:expzhao}, a much better agreement with the experimental data is achieved
when the the $^2$D resonances mediated by spin-orbit effects are taken into account in addition to
the R-Matrix cross section. For the generation of the incoherent sum of the  R-Matrix +
perturbation-theory rate coefficient shown in Fig.~\ref{fig:expzhao} we have used the $^2$D Auger
and radiative rates as calculated by Cowan's code. The resonance energy, however, has been
determined by using the experimentally observed energy splittings between the $^2$D resonances and
neighboring resonances obtained from R-Matrix theory. For example, the experiment yields
3p$^5$3d$^2$($^3$P)\,$^2$P and 3p$^5$3d$^2$($^3$F)\,$^2$D resonance energies of 14.8 and 15.3 eV,
respectively (Tab.~\ref{tab:fit}). The level difference equals 0.5 eV. Since our R-matrix
calculation yields a 3p$^5$3d$^2$($^3$P)\,$^2$P resonance energy of 16.4 eV, we chose 16.9 eV to be
the 3p$^5$3d$^2$($^3$F)\,$^2$D resonance energy.  Moreover, for the comparison in
Fig.~\ref{fig:expzhao} the theoretical merged-beams rate coefficient was multiplied by a factor
0.55 and shifted by 1.13~eV to lower energies. The difference between experimental and theoretical
cross sections of nearly a factor 2 is found to be independent of the inclusion of 4s, 4p and 4d
orbitals in the set of target states.

\subsection{Line shape of the 3p$^5$3d$^2$($^3$F)\,$^2$F resonance}\label{sec:3p53d2}

\begin{figure}[ttt]
\begin{center}
\includegraphics[width=\figurewidth]{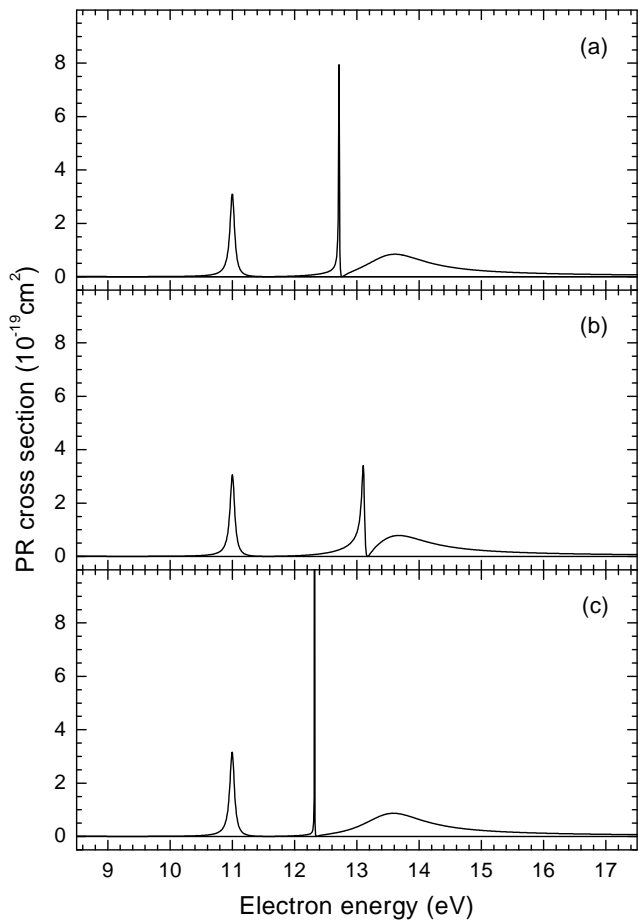}
\end{center}
\caption{ \label{fig:resres} Partial cross sections for recombination via $^2$F$^\mathrm{o}$ states
[in the order of increasing energy: 3p$^5$3d($^3$F)4s\,$^2$F, 3p$^5$3d($^1$F)4s\,$^2$F and
3p$^5$3d$^2$($^3$F)\,$^2$F] calculated in the R-matrix approach, showing the variation of the
resonance line shapes as a function of energy separation (a) as calculated ab initio, (b) 0.5
times, and (c) 1.5 times the separation in (a) with no other parameter varied.}
\end{figure}

Clearly, the theoretical line shape of the broad 3p$^5$3d$^2$($^3$F)\,$^2$F resonance is
asymmetric. As already mentioned in the introduction, Gorczyca \emph{et al.} \cite{Gorczyca1997}
have attributed this asymmetry to interference between RR and DR pathways.  However, this
interpretation may be doubted in view of the fact that in the relevant energy range the RR cross
section is orders of magnitude smaller than the DR cross section. Naively, one would expect that a
significant interference between DR and RR requires cross sections of similar  magnitude. It must
be kept in mind, however, that the asymmetry parameter $Q$ is also governed by the natural line
width $\Gamma_d$. Inspection of Eq.~\ref{eq:Q} shows that a large value of $\Gamma_d$ can make up
for a small RR transition matrix element $\langle j\vert R\vert f \rangle$. Fig.~\ref{fig:drrr}a
shows the RR, DR and interference contributions to the recombination cross section  in the energy
region of the 3p$^5$3d$^2$($^3$F)\,$^2$F resonance as calculated by perturbation theory. With the
width of the 3p$^5$3d$^2$($^3$F)\,$^2$F resonance due to its Super-Coster-Kronig decay channel
being extraordinarily large ($\Gamma_d = 1.06$~eV obtained from Cowan's atomic-structure code
\cite{Cowan1981}) the asymmetry in the merged-beams rate coefficient is retained even when the RR
cross section is artificially lowered by a factor of 10 (Fig.~\ref{fig:drrr}b). The theoretical
asymmetry parameter changes from $Q=3.97$ to $Q=12.6$. Our experimental findings of $\Gamma_d =
0.89$~eV and of $Q=6.3$ (see Tab.~\ref{tab:fit}) correspond to these theoretical results.

Apart from  interference between DR and RR another source of distorted line shapes is the
interaction between overlapping DR resonances of the same symmetry \cite{Griffin1994}. From our
R-Matrix calculation we find two further resonances of $^2$F symmetry, namely the resonances
3p$^5$3d($^3$F)4s\,$^2$F and 3p$^5$3d($^1$F)4s\,$^2$F located at 11.00 and 12.71 eV, respectively.
As shown in Fig.~\ref{fig:resres}a, especially the latter is overlapping with the broad
3p$^5$3d$^2$($^3$F)\,$^2$F resonance.  In Figs.~\ref{fig:resres}b and \ref{fig:resres}c we show how
the line shape of the 3p$^5$3d$^2$($^3$F)\,$^2$F resonance depends on its (artificially varied)
distance to the 3p$^5$3d($^1$F)4s\,$^2$F resonance. Clearly the shapes of both resonances are
affected by their mutual separation. The experimental observation of this effect, however, is
hindered by the presence of  the $^2$D resonances close to the 3p$^5$3d($^1$F)4s\,$^2$F resonance.
Under this circumstance the experimental resolution does presently not suffice to reveal the
theoretically predicted subtle resonance-resonance interference.

\section{Conclusions}\label{sec:conc}

In the experimental energy range 0-45~eV photorecombination of Sc$^{3+}$(3s$^2$3p$^6$) ions is
dominated by the 3p$^5$3d\,$^1$P\,$nl$ and 3p$^5$4s\,$^1$P\,$nl$ Rydberg series of DR resonances.
Significant differences to the recombination of isoelectronic Ti$^{4+}$ ions \cite{Schippers1998}
are found that are due to subtle differences in the electronic structure of Sc$^{3+}$ and
Ti$^{4+}$. Level energies, widths and resonance strengths were determined for a number of doubly
excited Sc$^{2+}$(3s$^2$3p$^5$3d$^2$) and Sc$^{2+}$(3s$^2$3p$^5$3d\,4s) states. Furthermore, we
find a small experimental evidence for an asymmetry of the 3p$^5$3d$^2$($^3$F)\,$^2$F resonance
line shape.  A more precise determination of the line shape from the experimental data would
require less statistical uncertainties especially in the high energy tail of the resonance. In the
present measurement the high background level from electron capture in Sc$^{3+}$ collisions with
residual gas molecules prevented this requirement from being fulfilled. Moreover, the peaks on the
low energy side of the 3p$^5$3d$^{2~2}$F resonance leave additional room for ambiguity.

Starting from rigorous continuum-bound transition theory the Sc$^{3+}$ photorecombination cross
section has been calculated using both the R-matrix method and a perturbative treatment. We found
that the line shape of the 3p$^5$3d$^2$($^3$F)\,$^2$F resonance is influenced by both interacting
resonances and interference between RR and DR. In spite of the fact that the Sc$^{3+}$ RR cross
section is almost negligibly small, DR-RR interference becomes noticeable due to the large
3p$^5$3d$^2$($^3$F)\,$^2$F resonance width of 0.89~eV. Moreover, we have shown that relativistic
effects play an important role in the recombination of Sc$^{3+}$. Three experimentally observed
strong 3p$^5$3d$^2$\,$^2$D and 3p$^5$3d4s\,$^2$D resonances can only be reproduced theoretically by
invoking relativistic effects. Our R-matrix calculations as well as the calculations of Gorczyca
\emph{et al.} \cite{Gorczyca1997} that both have been restricted to LS-coupling fail in reproducing
these strong experimental peaks.

Our results show the limits of the independent-processes and isolated-resonances approximations
usually made in photorecombination calculations. A more stringent test of our theoretical results
could in principle be provided by the measurement of high-resolution photoionization cross sections
of Sc$^{2+}$(3p$^6$3d$^{~2}$D) ions. Results of such an experiment will be reported elsewhere
\cite{Schippers2002}.

\begin{acknowledgments}

Technical support by the Heidelberg accelerator and TSR groups is gratefully acknowledged. The
experimental work has been supported by the Deutsche Forschungsgemeinschaft (DFG, grant no.\ Mu
1068/8). A.C and M.E.B were supported by a NATO collaborative grant (no.\ PST/CLG 976362) and L.B.Z
was supported by the National Natural Science Foundation through grant no.\ 19974006.

\end{acknowledgments}

\appendix

\section{Convolution of a Fano-profile with a Gaussian}

The convolution of a Fano profile (Eq.~\ref{eq:Fano}) with a normalized Gaussian
\begin{equation}\label{eq:gau}
G(E) = \frac{2}{\Delta E}\sqrt{\frac{\ln 2}{\pi}}\exp\left[-\frac{4(\ln2)\,E^2}{(\Delta E)
^2}\right]
\end{equation}
with $\Delta E$ being its full width at half maximum (FWHM), yields
\begin{eqnarray}
  C(E) &=& \int_{-\infty}^\infty F(E')\,G(E'-E)\,dE' \label{eq:c1}\\
   &=& A\frac{2}{Q^2\Gamma\sqrt{\pi}}\,\left[\frac{1}{\pi}\,\int_{-\infty}^\infty
   \frac{[Q y+(t-x)]^2}{(t-x)^2+y^2}e^{-t^2}dt-1\right]\nonumber\\
   &=& A \frac{2\sqrt{\ln 2}}{\Delta E\sqrt{\pi}}\,
  \left[\left(1\!-\!\frac{1}{Q^2}\right)\Re[w(z)]-\frac{2}{Q}\Im[w(z)]\right]\nonumber
\end{eqnarray}
where the definitions
\begin{eqnarray}
 t &=& \frac{2\sqrt{\ln 2}(E'-E)}{\Delta E}, \label{eq:tdef}\\
 x &=& \frac{2\sqrt{\ln 2}(E_d-E)}{\Delta E},\label{eq:xdef}\\
 y &=& \frac{\Gamma\sqrt{\ln 2}}{\Delta E},\label{eq:ydef}\\
 z &=& x+iy\label{eq:zdef}
\end{eqnarray}
have been used and where $w(z)$ denotes the complex error function \cite{Abramowitz1964}. Its real
and imaginary parts can efficiently be calculated with a numerical algorithm by Humlicek
\cite{Humlicek1979}. It should be noted that for $Q\to\infty$ Eq.~\ref{eq:c1} yields the Voigt
profile, i.\,e.\ the convolution of a Lorentzian (Eq.~\ref{eq:lor}) with a Gaussian.


\begin{thebibliography}{XX}

\bibitem{Alber1984}
G. Alber, J. Cooper, and A.~R.~P. Rau, Phys. Rev. A {\bf 30},  2845  (1984).

\bibitem{Badnell1992}
N.~R. Badnell and M.~S. Pindzola, Phys. Rev. A {\bf 45},  2820  (1992).

\bibitem{Zimmermann1997}
M. Zimmermann, N. Gr\"{u}n, and W. Scheid, J. Phys. B {\bf 30},  5259  (1997).

\bibitem{Behar2000b}
E. Behar, V.~L. Jacobs, J. Oreg, A. Bar-Shalom, and S.~L. Haan, Phys. Rev. A
  {\bf 62},  030501  (2000).

\bibitem{Knapp1995a}
D.~A. Knapp, P. Beiersdorfer, M.~H. Chen, J.~H. Scofield, and D. Schneider,
  Phys. Rev. Lett. {\bf 74},  54  (1995).

\bibitem{Gorczyca1997}
T.~W. Gorczyca, M.~S. Pindzola, F. Robicheaux, and N.~R. Badnell, Phys. Rev. A
  {\bf 56},  4742  (1997).

\bibitem{Schippers1999a}
S. Schippers, T. Bartsch, C. Brandau, A. M\"{u}ller, J. Linkemann, A.~A.
  Saghiri, and A. Wolf, Phys. Rev. A {\bf 59},  3092  (1999).

\bibitem{Cowan1981}
R.~D. Cowan, {\em The Theory of Atomic Structure and Spectra} (University of
  California Press, Berkeley, 1981).

\bibitem{Hansen1996}
J.~E. Hansen and P. Quinet, J. Elec. Spectros. Rel. Phenom. {\bf 79},  307
  (1996).

\bibitem{Mueller2001}
A. M\"{u}ller, S. Schippers, A.~M. Covington, A. Aguilar, G. Hinojosa, R.~A.
  Phaneuf, M.~M. Sant'Anna, A.~S. Schlachter, J.~D. Bozek, and C. Cisneros,  in
  {\em {XXII} International Conference on Photonic, Electronic, and Atomic
  Collisions, Santa Fe, New Mexico, {USA}, July 18-24, 2001, Abstracts of
  Contributed Papers}, edited by S. Datz, M.~E. Bannister, H.~F. Krause, L.~H.
  Saddiq, D. Schultz, and C.~R. Vane (Rinton Press, Princeton, New Jersey,
  2001), p.\ 52.

\bibitem{Zhao2000}
L.~B. Zhao, A. Ichihara, and T. Shirai, Phys. Rev. A {\bf 62},  022706  (2000).

\bibitem{Kilgus1992}
G. Kilgus, D. Habs, D. Schwalm, A. Wolf, N.~R. Badnell, and A. M\"{u}ller,
  Phys. Rev. A {\bf 46},  5730  (1992).

\bibitem{Lampert1996}
A. Lampert, A. Wolf, D. Habs, J. Kenntner, G. Kilgus, D. Schwalm, M.~S.
  Pindzola, and N.~R. Badnell, Phys. Rev. A {\bf 53},  1413  (1996).

\bibitem{Schippers2000b}
S. Schippers, T. Bartsch, C. Brandau, A. M\"{u}ller, G. Gwinner, G. Wissler, M.
  Beutelspacher, M. Grieser, A. Wolf, and R.~A. Phaneuf, Phys. Rev. A {\bf 62},
   022708  (2000).

\bibitem{Schippers2001c}
S. Schippers, A. M\"{u}ller, G. Gwinner, J. Linkemann, A.~A. Saghiri, and A.
  Wolf, Astrophys. J. {\bf 555},  1027  (2001).

\bibitem{Grieser1991}
M. Grieser, M. Blum, D. Habs, R.~V. Hahn, B. Hochadel, E. Jaeschke, C.~M.
  Kleffner, M. Stampfer, M. Steck, and A. Noda,  in {\em Proceedings of the
  19th International Symposium on Cooler Rings and Their Applications, Tokyo,
  Japan, November 5 -- 8, 1990}, edited by T. Katayama and A. Noda (World
  Scientific, Singapore, 1991), pp.\ 190--198.

\bibitem{Hochadel1994a}
B. Hochadel, F. Albrecht, M. Grieser, D. Habs, D. Schwalm, E. Szmola, and A.
  Wolf, Nucl. Instrum. Methods A {\bf 343},  401  (1994).

\bibitem{Rinn1982}
K. Rinn, A. M\"{u}ller, H. Eichenauer, and E. Salzborn, Rev. Sci. Instrum. {\bf
  53},  829  (1982).

\bibitem{Schlachter1983}
A.~S. Schlachter, J.~W. Stearns, W.~G. Graham, K.~H. Berkner, R.~V. Pyle, and
  J.~A. Tanis, Phys. Rev. A {\bf 27},  3372  (1983).

\bibitem{Davies1969}
P.~C.~W. Davies and M.~J. Seaton, J. Phys. B {\bf 2},  757  (1969).

\bibitem{Clementi1974}
E. Clementi and C. Roetii, At. Data Nuc. Data Tab. {\bf 14},  177  (1974).

\bibitem{Hibbert1975}
A. Hibbert, Comput. Phys. Commun. {\bf 9},  141  (1975).

\bibitem{Sobelman}
I. Sobleman, \emph{Atomic Spectra and Radiative Transitions} (Springer, Berlin, 1992).

\bibitem{Fano1961}
U. Fano, Phys. Rev. {\bf 124},  1866  (1961).

\bibitem{Mitnik1999}
D.~M. Mitnik, M.~S. Pindzola, and N.~R. Badnell, Phys. Rev. A {\bf 59},  3592
  (1999).

\bibitem{Schippers1998}
S. Schippers, T. Bartsch, C. Brandau, G. Gwinner, J. Linkemann, A. M\"{u}ller,
  A.~A. Saghiri, and A. Wolf, J. Phys. B {\bf 31},  4873  (1998).

\bibitem{Burgess1964b}
A. Burgess, Mem. Roy. Astron. Soc. {\bf 69},  1  (1964).


\bibitem{Gorczyca1998a}
T.~W. Gorczyca, private communication.


\bibitem{Tiwary1983}
S.~N. Tiwary, A.~E. Kingston, and A. Hibbert, J. Phys. B {\bf 16},  2457
  (1983).

\bibitem{Altun1999}
Z. Altun and T. Manson, J. Phys. B {\bf 32},  L255  (1999).

\bibitem{Zhao2001}
L.~B. Zhao and T. Shirai, Phys. Rev. A {\bf 63},  010703(R)  (2001).

\bibitem{Griffin1994}
D.~C. Griffin, M.~S. Pindzola, F. Robicheaux, T.~W. Gorczyca, and N.~R.
  Badnell, Phys. Rev. Lett. {\bf 72},  3491  (1994).

\bibitem{Schippers2002}
S. Schippers, A. M\"{u}ller, S. Ricz, M.~E. Bannister, G.~H. Dunn, J. Bozek, A.~S. Schlachter,  G.
Hinojosa, C. Cisneros,  A. Aguilar, A. Covington, M. Gharaibeh, and R.~A. Phaneuf, to be published.

\bibitem{Abramowitz1964}
M. Abramowitz and I.~A. Stegun, {\em Handbook of Mathematical Functions} (Dover
  Publications, New York, 1964), chapter 7.

\bibitem{Humlicek1979}
J. Humlicek, J. Quant. Spectrosc. Radiat. Transfer {\bf 21},  309  (1979).

\bibitem{Martin1999}
W.~C. Martin, J. Sugar, A. Musgrove, W.~L. Wiese, J.~R. Fuhr, D.~E. Kelleher,
  K. Olsen, P.~J. Mohr, G. Dalton, M. Douma, R. Dragoset, S. Kotochigova, L.
  Podobedova, E. Saloman, C. Sansonetti, and G. Wiersma, {\em {NIST} Atomic
  Spectra Data Base}, 2. ed., National Institute of Standards and Technology,
  Gaithersburg, Maryland 20899-3460, USA, 1999,
  {\tt http:$//$physics.nist.gov$/$cgi-bin$/$AtData$/$main$\_$asd}.
\end{thebibliography}
\end{document}